# Hybrid Superscattering Driven by Toroidal Dipole

Kislov[1,2,3,*] D., Borovkov[1] D., Huang[3] L., Kuznetsov[1] A., Canos Valero[4] A., Ipatovs[5] A., Bobrovs[5] V., Fedotov[2,*] V., Gao[6] L., Xie[3] S., Xu[7] Y., Luo[8] J., Baranov[1] D., Arsenin[1] A., Bolshakov[1] A., Shalin[1,2,3,6,*] A.S.

[1]Center for Photonics and 2D Materials, Moscow Institute of Physics and Technology, Dolgoprudny 141700, Russia
[2]Centre for Photonic Science and Engineering, Skolkovo Institute of Science and Technology, Moscow, 121205, Russia.
[3]Beijing Institute of Technology Beijing 100081, China
[4]Institute of Physics, University of Graz, and NAWI Graz, 8010, Graz, Austria
[5] Riga Technical University, Institute of Photonics, Electronics and Telecommunications, 1048 Riga, Latvia
[6]School of Optical and Electronic Information, Suzhou City University, Suzhou 215104, China.
[7]Guangdong University of Technology, Guangzhou 510006, China
[8] School of Physical Science and Technology, Soochow University, Suzhou 215006, China

*Corresponding Author E-Mail: alexandesh@gmail.com; denis.a.kislov@gmail.com; V.Fedotov@skoltech.ru

## Abstract

The dynamic toroidal dipole is a peculiar elementary source of electromagnetic radiation, which cannot be represented in terms of the familiar electric and magnetic multipoles. Since its experimental demonstration some 15 years ago, toroidal dipole has been the subject of ever-growing interest both theoretically and experimentally. Most of the research efforts so far have been focused on enhancing the electromagnetic signatures of the toroidal dipole (and higher-order toroidal multipoles), which is known to couple weakly to free space.

Here we report on a surprising finding that the toroidal dipole can, in fact, be engaged in the enhancement of electromagnetic scattering per se driving the so-called superscattering—the regime of anomalously strong light scattering where the total cross-section of the effect exceeds the fundamental single-channel limit. We introduce a new paradigm of hybrid superscattering enabled by the toroidal dipole, which we implement with a dielectric scatterer of a simple geometry, and demonstrate for the first time that two complementary mechanisms of superscattering—the Friedrich-Wintgen mechanism and resonance overlap—can act synergistically to yield the substantially enhanced effect.

Using coupled-dipole theory, full-wave numerical modeling and coupled-mode theory, we identify and quantify the dominant multipolar contributions and show that the normalized scattering cross-section exceeds the dipole limit due to a toroidal dipole-magnetic quadrupole interplay. These findings are supported by experimental measurements in the GHz frequency range using a dimer of ceramic cubes, which confirm both the spectral and spatial features of toroidal superscattering.

Our results open a new powerful route to engineering strong light-matter interaction via peculiar toroidal modes (never observed before) with potential applications in toroidal superscattering metamaterials and metasurfaces, photonic devices, and sensors.

## Introduction

Dynamic toroidal multipoles are regarded as a separate multipole family [1–3] or a part of so-called "exact" multipoles [4]. They were first extracted from the multipole expansion and theoretically analyzed by Afanasiev and Dubovik in their seminal works [5,6]. The most primitive member of dynamic toroidal multipoles – the toroidal dipole (TD) – can be represented by poloidal currents oscillating along the meridians of an imaginary torus or, alternatively, by a vortex oscillating magnetic dipole moments (Fig.1a). While TD has the same parity and radiation pattern as an electric dipole (ED) [2], there are some important fundamental differences between these two types of elementary radiation sources. In particular, TD is a purely current source [6]; it interacts with $d\mathbf{E}/dt$ rather than $\mathbf{E}$ [7], the field of vector potential it generates is different from that produced by ED and the difference cannot be eliminated by gauge transformations [8]. As the result, engaging TD (and higher toroidal multipoles, in general) has enabled the demonstration of several intriguing scattering regimes and optical effects, most notably, a new mechanism of electromagnetic transparency and the non-trivial non-radiating charge-current configuration [8,9] – the so-called dynamic anapole [10,11] with the controversy surrounding the non-trivial vector potential field yet to be resolved [12], hybrid anapoles [13,14], a new type of optical activity [15], an anapole nanolaser [16], and the resonant Kerker effect [4,17–20]. More recently, the excitation of TD has been shown to underpin a new truly 'toroidal' regime of permittivity sensing [21] and predator-prey nonreciprocal interaction [22].

Due to its 'vortex-like' symmetry, however, the toroidal dipole mode couples poorly to external radiation fields and exhibits weak electromagnetic scattering [23]. Thus, the peculiar phenomena associated with TD (and higher-order toroidal multipoles) are often difficult to observe let alone exploit, as they are masked by the effects due to the conventional multipoles. The toroidal response is currently distilled using the following two mainstream strategies. The first involves the engineering of sub-wavelength metallic scatterers and antennae of explicit toroidal symmetry, which enables one to suppress the excitation of the conventional multipoles [9,24–28]. The second favours the operation beyond the long-wavelength limit in optically large dielectric structures [18,29–32] and allows relative enhancement of TD scattering, since the electromagnetic fields radiated by TD scale as $(a/\lambda)^3$ rather than $(a/\lambda)$ and $(a/\lambda)^2$, as in the case of ED and magnetic dipole (MD) scattering, respectively [1,3]. While those strategies have proven to be effective in exposing weak electromagnetic signatures of toroidal multipoles [33], the idea of engaging toroidal modes to enhance the efficiency of electromagnetic scattering as such and, in particular, demonstrate the effect of so-called superscattering seems counterintuitive and, so far, has not been explored.

Superscattering is a special regime of enhanced scattering, which is characterized by the cross-section of the effect being many times greater than the so-called "single-channel limit" [34]. The latter dictates that the maximal efficiency of scattering by a spherical nanoparticle for each multipole resonance (i.e., scattering channel) depends only on the wavelength of the resonance, $\lambda$, and the angular momentum of the relevant multipole, $l$: $\sigma_0^l = (2l+1)\lambda^2/2\pi$. For instance, for a dipole resonance ($l = 1$) the single-channel limit is equal to $\sigma_0 = 3\lambda^2/2\pi$, and so *superscattering* occurs when the actual scattering cross-section, $\sigma_{\text{sca}}$, exceeds $\sigma_0$. Superscattering is normally achieved by combining different multiple resonances at the same wavelength [34–38]. Even though the single-channel limit applies to scattering through each multipole channel (at least in the spherical geometry), several multipole resonances can be made to spectrally overlap by modifying the properties of a nanoparticle. As the result, the total scattering of the nanoparticle will dramatically increase and can in principle exceed the single-channel limit. There is also another way of engineering superscattering, which has been recently demonstrated [39,40] based on the Friedrich-Wintgen mechanism [41]. The latter, while normally responsible for the appearance of so-called bound states in the continuum (BICs) and quasi-BICs with ultrahigh Q-factors (electromagnetic modes exhibiting vanishing or suppressed scattering losses) [42–44], may also lead to the constructive interference of several modes within one scattering channel, thus, enabling superscattering (super-multipole) regime bypassing the single-channel limit.

In this work we show that not only TD opens new horizons in controlling light scattering but also expands the very concept of superscattering. In particular, we extend the Friedrich-Wintgen mechanism to the case of TD scattering and confirm that superscattering can be achieved even with a pure TD under plane-wave illumination. Moreover, by spectrally overlapping TD and magnetic quadrupole resonances we managed to combine the Friedrich-Wintgen mechanism with the common approach to superscattering, and for the first time introduce the regime of hybrid superscattering with its cross-section substantially exceeding the single-channel limit. Below we describe the conditions for such toroidal superscattering regime to occur, as well as demonstrate it numerically and experimentally in a simple dielectric structure.

**Initial Design of the System**

In this paper we will distinguish among the *exact ED* moment, $\mathbf{p}$, *basic or primitive ED* moment, $\mathbf{d}$, and TD moment, $\mathbf{T}$, which are interconnected in the following way [4]:

$$\mathbf{p} = \mathbf{d} + \mathbf{T} + \text{Higher order toroidal terms (mean-square radii)}. \tag{1}$$

While TD moment is commonly supplemented with the factor $ik/c$, we elected to include it in the definition of the moment.

As a starting point we need to design a scatterer, which can support the excitation of TD under plane-wave illumination with the latter providing a dominant scattering channel. Since TD may be represented by a vortex of MDs [28], two MDs oscillating with the same amplitude but opposite phases and placed some distance apart should give rise to non-zero TD moment. It is important to recall at this point that TD and the basic ED have the same parity and identical radiation patterns and, hence, can interfere with each other. We aim to minimize the interference of the two in order to isolate the contribution of TD to scattering, which is only possible if the basic ED has a vanishing moment.

We argue that a nearly pure TD scatterer can be formed by a pair of Si spheres under plane-wave illumination (Fig.1d) provided the following conditions are satisfied: (i) both spheres exhibit an MD resonance at the same wavelength, (ii) MD moments induced in the spheres oscillate with a relative phase difference close to $\pi/2$ (iii) the basic ED moment of the overall structure is negligible (i.e., the exact ED moment contains TD contribution only), (iv) the wave-vector of the incident wave should be parallel to the axis of the dimer. The easiest way to satisfy the first condition is to ensure that the spheres are identical (hereinafter we will consider the spheres with the parameters from Fig. 1 caption). Conditions (ii) and (iii) can be satisfied by adjusting such geometric parameters of the composite scatterer as the distance between the spheres and their radius.

As evident from the plot in Fig. 1e, the contributions of quadrupoles and higher-order multipoles to scattering by isolated spheres near their MD resonance are negligible. Therefore, the optical response of the composite scatterer can be described in a simple manner using the coupled-dipole approximation [45,46]. The analytics devoted to finding the exact electric $\mathbf{p}_j$ and magnetic $\mathbf{m}_j$ ($j$ is the particle index) dipole moments of both particles with taking into account coupling between the particles is given in the Supplementary Information S1. Once $\mathbf{p}_j$ and $\mathbf{m}_j$ have been obtained, the extinction cross-section of the composite scatterer (dimer) can be calculated using the following expression (Eq.7 in [45]):

$$\sigma_{ext} = \frac{k_d}{\varepsilon_0 \varepsilon_d |E_0|^2} \operatorname{Im} \sum_{j=1}^{N} \left[ \mathbf{E}_j^{0*} \mathbf{p}_j + \mu_0 \mathbf{H}_j^{0*} \mathbf{m}_j \right].$$ (2)

To aid the analysis of the effect we have further simplified our theoretical model by neglecting dissipation in Si. That rendered the particles non-absorbing and eliminated the distinction between their scattering and extinction cross-sections $\sigma_{sca} = \sigma_{ext} = \sigma$. (which we will refer to below as simply 'cross-section').

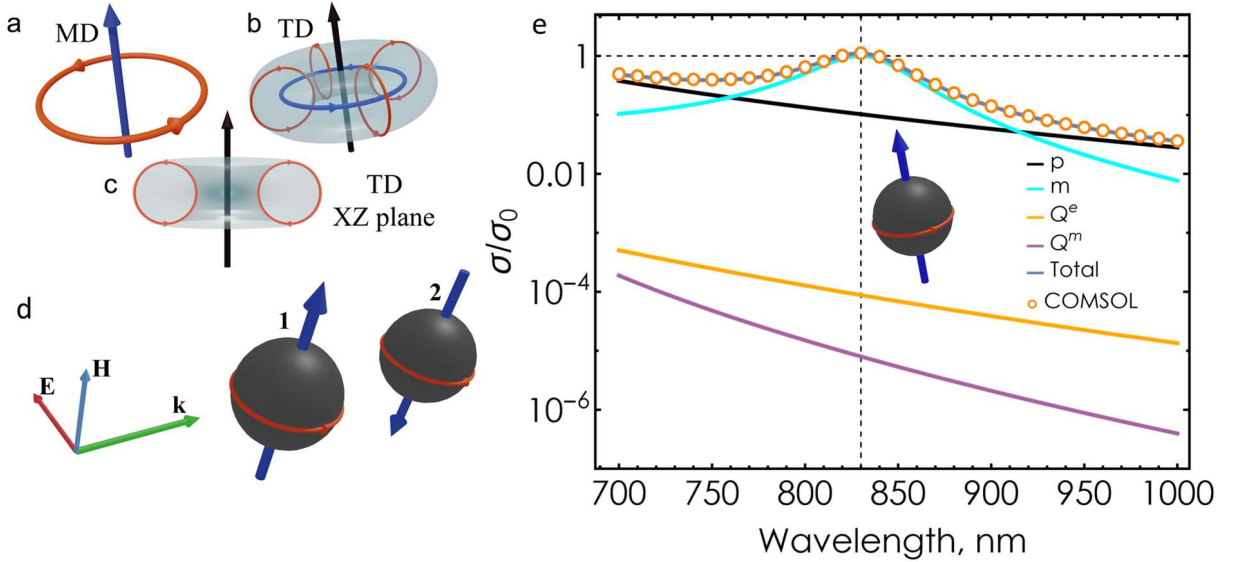

**Figure 1.** a) Schematic representation of a magnetic dipole. b) Schematic representation of a toroidal dipole. c) XZ cross-section of a toroidal dipole. d) Illustration of a dimer formed by two Si spheres supporting MDs and illuminated by a plane wave along its axis. e) Mie multipole contributions to the normalized scattering cross-section, $\sigma/\sigma_0$, of a single sphere with a radius of 100 nm and refractive index $n = 4$. Here, $\mathbf{p}$ denotes the normalized contribution of the exact ED $\mathbf{m}$ - exact MD and $\mathbf{Q^{e,m}}$ - exact electric and magnetic quadrupoles, respectively. Since electric and magnetic quadrupoles of the sphere are negligible, the total scattering cross-section can be described just with the exact electric and magnetic dipoles.

## Results and Discussions

### X.1. Toroidal Superscattering

First, we perform the multipole decomposition for a single isolated sphere to determine the spectral location of its MD resonance under plane-wave illumination (Fig. 1e). The sphere has the radius $r = 100$ nm and the refractive index of the material it is made of $n = 4$, which corresponds roughly to lossless Si. As evident from Fig. 1e, scattering by the sphere via MD resonance peaks at the wavelength $\lambda \sim 830$nm, and the value of the scattering cross-section normalized to the single-channel limit, as expected, reaches 1. A small contribution from the exact ED lifts the total cross-section of the sphere slightly above 1 (Fig.1e).

Next, we introduce a second sphere such that the axis of the resulting dimer is parallel to the direction of the incident wave. For the range of wavelengths corresponding to MD resonance of a single sphere we use Eq. (2) to study the variation of the normalized cross-section of the dimer while changing the distance between the centres of the two spheres, $R$. At long distances the interaction between the particles is weak and the cross-section of the complex scatterer is seen to oscillate depending on the phase difference between MD moments of the spheres (see Fig. S2.1 in Supplementary Information S2). However, when particles are close enough the interaction between them starts to play a crucial role. In particular, when the spheres nearly touch each other the normalized cross-section of the dimer, $\sigma/\sigma_0$, reaches the staggering value of 2.7 which is much higher than just the sum of two normalised cross-sections of non-interacting particles.

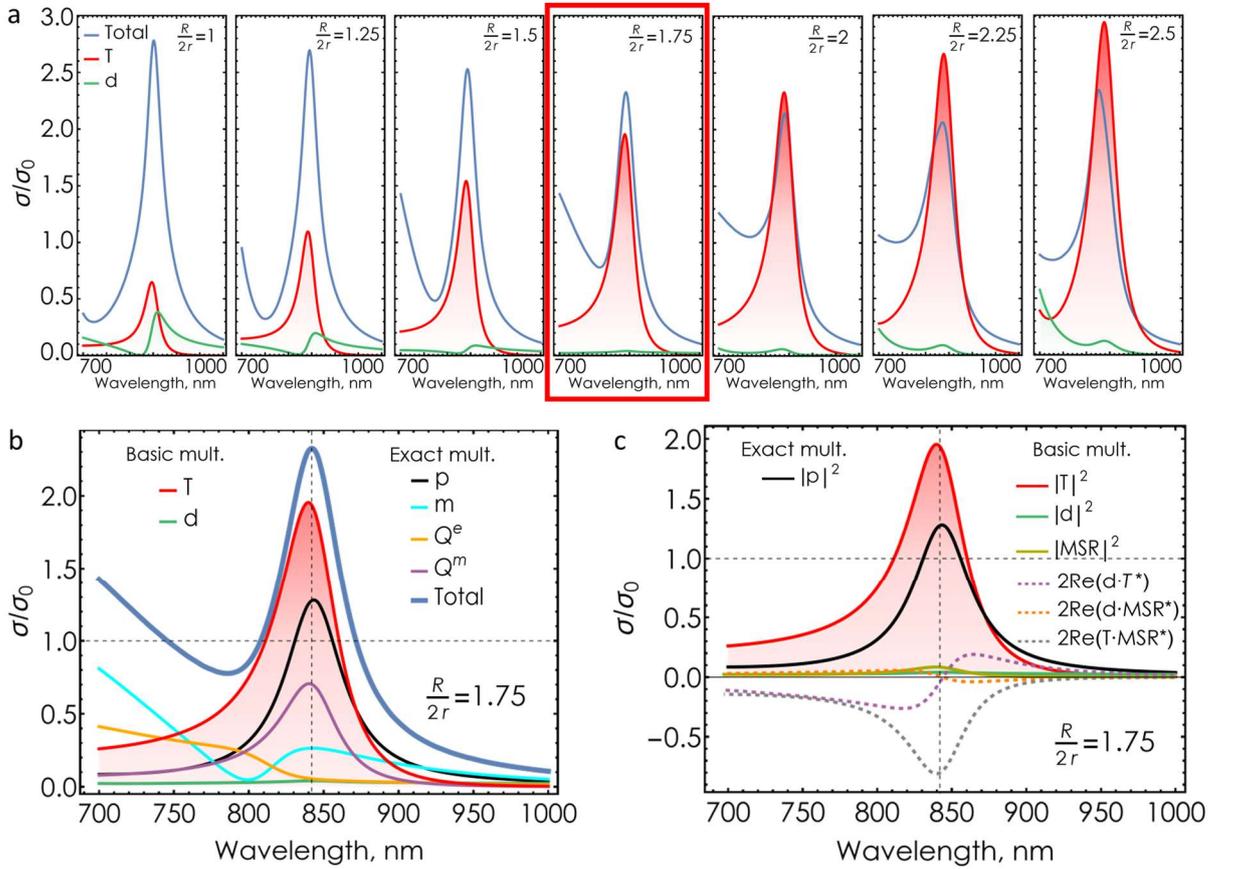

**Figure 2.** a) A set of plots comparing the normalized contributions of the basic ED (**d**, green line) and TD (**T**, red line) of the entire dimer to its total normalized cross-section (blue line) for selected interparticle separations. The case of nearly zero basic ED is marked with a red rectangle, and corresponds to $R/2r = 1.75$. MD contribution here is small. b) Same as the plot marked with a red rectangle in Fig.2a but showing also contributions from other multipoles. c) Contribution of various components of the exact ED (**p**) to the radiated power as a function of the excitation wavelength. Numerical simulations were carried out in COMSOL Multiphysics.

To find the regime of pure TD superscattering we modelled numerically the scattering of light by the dimer for a range of separations between the constituent particles, and performed the multipole decomposition for the composite scatterer in each case. Fig.2a represents the normalized contributions of the basic ED (**d**) and TD (**T**) to scattering by the dimer for $R$ in the rage between $2r$ and $5r$. As we mentioned above, one of the conditions for achieving pure TD superscattering is to ensure that the basic ED moment is close to zero. Fig.2a clearly shows that to a large extent this condition is satisfied when $R/2r = 1.75$ (i.e., $R = 350$nm). For this separation the normalized scattering cross-section of the dimer reaches 2.3 and the main contribution to scattering comes from TD.

The complete multipole decomposition in this case is shown in Fig.2b. Here, the normalized contribution of the exact ED, **p**, is smaller than that of TD alone even though the contribution of the basic ED, **d**, is close to zero. This is due to constructive interference of TD with its mean square radii (**MSR**) [30,47] (Eq. (1)). The power radiated by an exact ED (**p**) can be written:

$$P_{dip}^{rad} \propto |\mathbf{d} + \mathbf{T} + \mathbf{MSR}|^2 =$$

$$= \underbrace{|\mathbf{d}|^2}_{\rightarrow 0} + |\mathbf{T}|^2 + |\mathbf{MSR}|^2 + 2\left[ \underbrace{\mathrm{Re}\left(\mathbf{d} \cdot \mathbf{T}^*\right)}_{\rightarrow 0} + \underbrace{\mathrm{Re}\left(\mathbf{d} \cdot \mathbf{MSR}^*\right)}_{\rightarrow 0} + \mathrm{Re}\left(\mathbf{T} \cdot \mathbf{MSR}^*\right) \right] \qquad (3)$$

The contribution of various terms in Eq. (3) is shown in Fig. 2c. Evidently, due to vanishing value of the basic electric dipole moment, **d**, the main contributor to radiation is interference between the TD and MSR. The negative value of the interference term, $\mathrm{Re}\left(\mathbf{T} \cdot \mathbf{MSR}^*\right)$, makes the scattering contribution of an exact ED (**p**) less than that of a pure TD. Nevertheless, the normalized cross-section of the exact ED at the resonance wavelength is still noticeably higher than the single-channel limit, which is the manifestation of the unique TD-driven superscattering regime demonstrated for the first time. It is important to point out here that the excitation of a magnetic quadrupole (MQ) is also resonant at around 840nm and provides the second strongest contribution to superscattering.

## X.2. Explanations and Physical Mechanisms

To explain the formation of the toroidal response and the associated mechanism of super-scattering, we analyse the connection between the multipoles of the constituent particles and the multipoles of the whole dimer. To this end we plot in Fig. 3a the normalized scattering cross-section of the dimer and every non-zero term in Eq. (2), namely the MD and the exact ED induced in each sphere. It can be clearly seen that the normalized contributions of MDs in the two particles spectrally overlap and have the same magnitudes (being equal to 1) at the resonance wavelength. Moreover, the phase difference between them is close to $\pi$, which indicates that the dipole moments oscillate in anti-phase (Fig.3b). Note that the sum of the scattering cross-sections of all four dipole components is in a good agreement with the total cross-sections of the whole dimer calculated in the centre of mass of the structure using COMSOL Multiphysics (orange circles, Fig. 3a) and coupled-dipole approximation (brown curve Fig. 3a). In Figs. 3c and 3d we show the near-field distributions of the electric field induced around the spheres at the resonance. Clearly, the field distributions are characteristic of two MD modes with oppositely directed dipole moments, which results in predominantly toroidal dipolar response of the dimer. This is also evident from Fig. 2b, where we compare the strength of the resulting TD with other multipoles in terms of the normalized scattering cross-section.

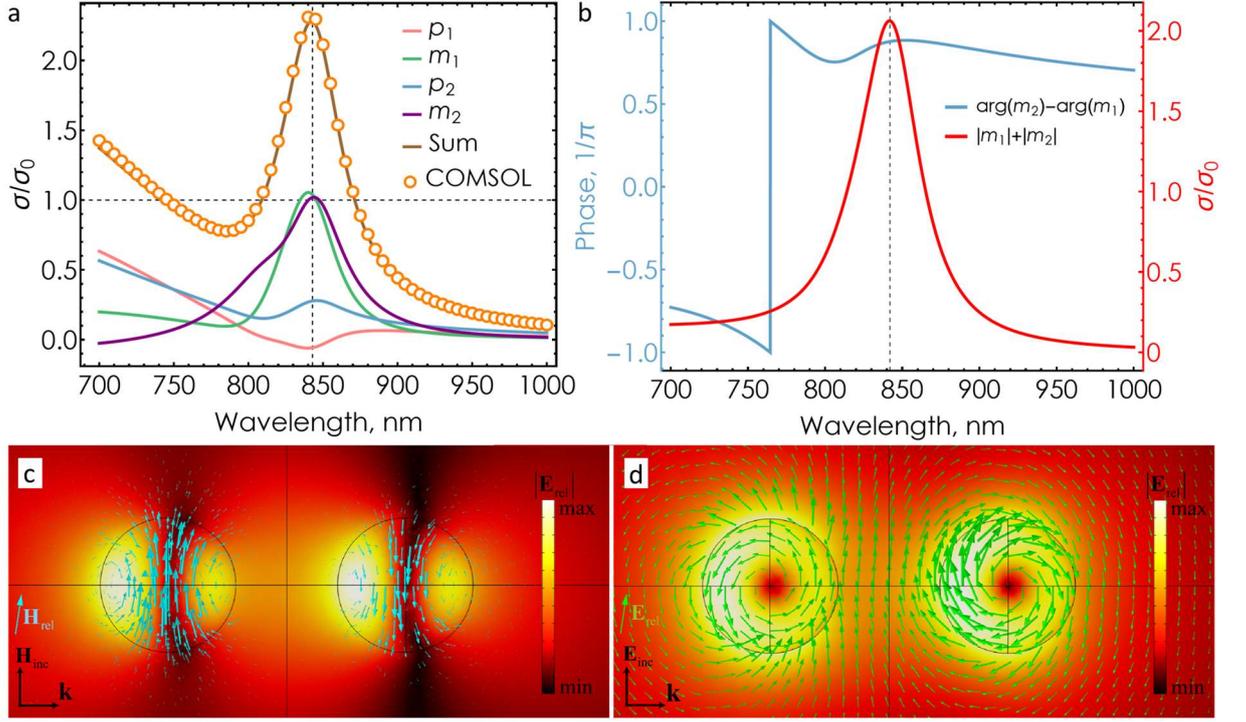

**Figure 3.** a) $\mathbf{p}_{1,2}, \mathbf{m}_{1,2}$ are the contributions of the exact electric and magnetic dipoles of first and second particles to the normalized scattering cross-section, calculated with coupled-dipole approximation at the points of *their* centers (Eq.**Ошибка! Источник ссылки не найден.**), orange circles – total cross-section numerically calculated in the center *of the system* for the model in COMSOL Multiphysics. b) The difference of phases of MDs of the spheres (blue line) and the sum of the amplitudes of the MDs calculated analytically (Eq.(2)) (red line). c,d) For the resonant wavelength $\lambda = 843$ nm the absolute value of the relative electric field is shown with color, while c) blue arrows show the direction of the relative magnetic field and d) green arrows – the direction of the relative electric field. It is clear that both spheres behave as MDs.

At this point it is important to recall that the combination of two MDs oscillating in anti-phase also gives rise to a magnetic quadrupole (MQ). While in our case MQ is about 3 times weaker than TD in terms of the normalized scattering cross-section, it still provides the second strongest contribution to overall scattering by the dimer at 830 nm (see Fig. 2b). The appearance of MD and MQ in a dimer can easily be shown analytically by expressing their moments in terms of point magnetic dipole moments (PMDs). Indeed, the total current density corresponding to a set of PMDs can be written as:

$$\mathbf{J} = \sum_j \mathbf{J}^j = \sum_j \nabla \times \left( \mathbf{m}^j \delta(\mathbf{r} - \mathbf{r}^j) \right) \quad , \tag{4}$$

here the summation is performed over all point magnetic dipole moments of the system. To express TD and MQ moments in terms of PMDs we substitute Eq. (4) into the expressions for TD and MQ moments (see Supplementary Information S3). After some algebraic transformations (for more details, see Supplementary Information S4) the expression for TD moment acquires the form:

$$\mathbf{T} = -\frac{1}{i\omega} \frac{k^2}{2} \left[ \sum_j \left[ \mathbf{r}^j \times \mathbf{m}^j \right] \right] \quad , \tag{5}$$

where $\mathbf{r}^j$ is the radius vector indicating the location of the j-th point magnetic dipole. Similarly, the Cartesian components of MQ moment take the form:

$$Q_{\alpha\beta}^m = \sum_s 3\left(r_\alpha^s m_\beta^s + r_\beta^s m_\alpha^s\right) - 2\left(\mathbf{m}^s \cdot \mathbf{r}^s\right)\delta_{\alpha\beta}, \qquad (6)$$

where $\alpha, \beta \in \{x, y, z\}$. Figure S4.1 (see Supplementary Information S4) shows the comparison of the normalized scattering cross-sections of TD and MQ extracted from full-wave numerical simulation and calculated using Eqs. (5) and (6). It is evident that the results produced by the analytical model and the exact numerical simulation are in a very good agreement confirming that the main contributions to the dimer scattering at the resonance are made by TD and MQ.

Clearly, the spectral overlap of TD and MQ resonances contributes to the observed super-scattering regime, similarly to the conventional mechanism of super-scattering [34,35,38,48]. I, not only the total intensity of TD and MQ channels exceeds the dipole limit. One can also see that $\sigma_p$ alone exceeds the dipole limit, which is not possible via the trivial resonance overlap. This suggests that another mechanism of super-scattering must be at play. Indeed, it is well known that the scattering cross-section of a multipole may exceed the respective single-channel limit if the scatterer lacks rotational and/or reflection symmetry [39].

To reveal the underlying mechanism of the observed behavior we perform the analysis of the modal spectrum of the system. The quasi-normal modes (QNMs) of a single spherically symmetric particle are represented by pure multipolar TE- and TM-polarized fields of angular momentum $\ell = 1,2,...$ (see Fig S5.1. in Supplementary Information S5). Arranging two particles in a dimer breaks spherical symmetry, thus, allowing new hybrid QNMs to couple to a few scattering channels at once. Usually, coupling between modes of a given scatterer radiating predominantly into the same scattering channel results in the emergence of a quasi-BIC state due to destructive interference of the modes via the Friedrich-Wintgen mechanism [43,49–53]. However, if the two original modes instead radiate into *different* scattering channels and interact non-radiatively, they may give rise to a super-scattering regime via the multipole mixing mechanism [39,54].

Fig. 4a shows the eigenfrequencies of the dimer as a function of center-to-center distance $R$. Coupling between two nanoparticles leads to the emergence of four hybrid modes (when considering the modes originating from the initial MD resonance of the sphere see Fig.S5.1). Two of them have azimuthal symmetry around the dimer's axis, $m = 0$. These are longitudinal anti-symmetric and symmetric hybrid QNMs formed by MDs oscillating in the particles in- and out-of-phase, respectively, along the axis of the dimer. Clearly, these modes cannot be excited by a plane wave propagating along the axis of the dimer, and thus are irrelevant for our analysis (more details on the distribution of electric and magnetic fields of eigenmodes can be found in Supplementary Information S6).

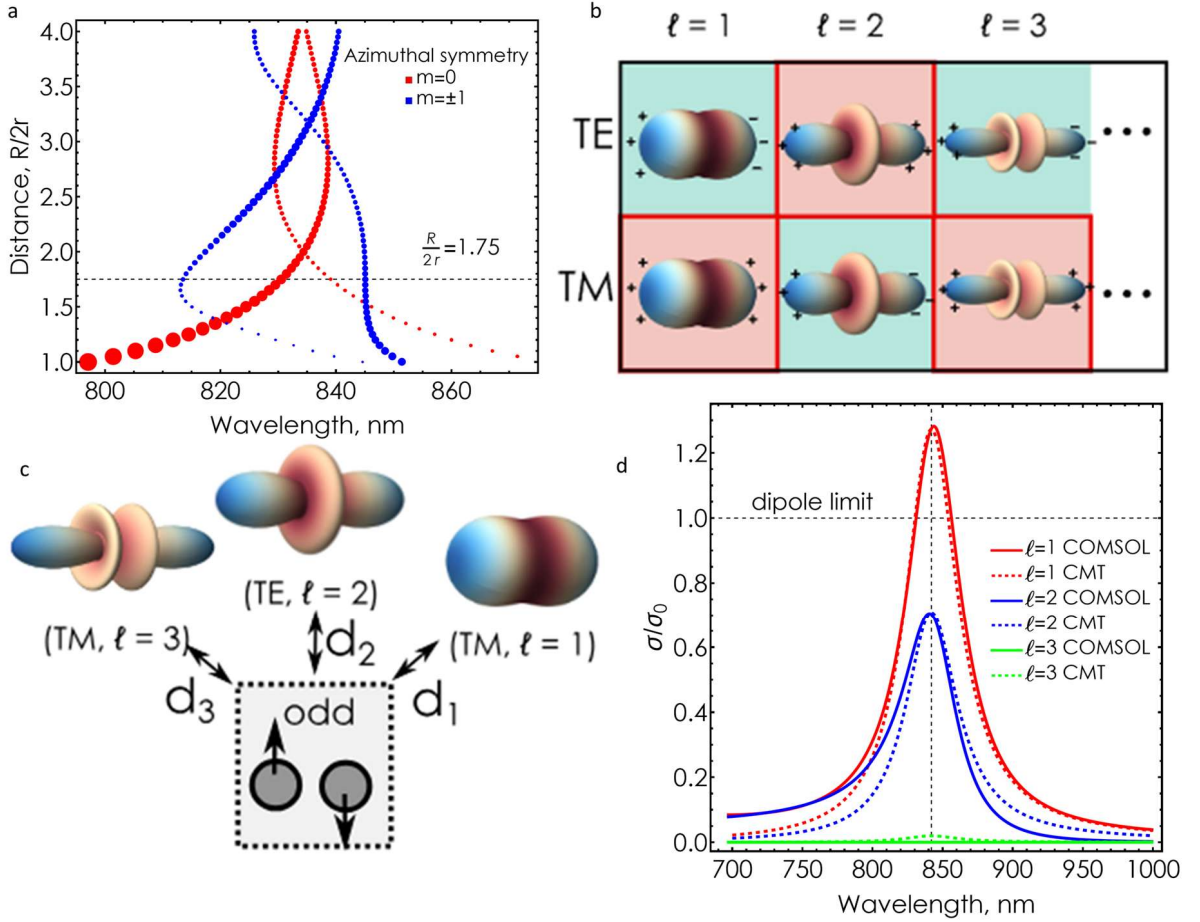

**Fig.4.** Modal picture of toroidal dipole super-scattering. (a) Complex-valued eigenfrequencies of a dimer of two dielectric $n=4$ particles of radius $r$ at a center-to-center distance $R$ in the spectral range of interest close to the fundamental MD resonance of the single particle. The size of the dots corresponds to the Q- factor of the mods. (b) The modal picture of the dimer in the context of the observed super-scattering behavior. Symmetric and anti-symmetric modes couple to an infinite ladder of TE- and TM-polarized $\ell=1,2,...$ scattering channels. (c) Coupled-mode picture of toroidal super-scattering by the dielectric dimer. The minimal model contains a single mode coupled to TM $\ell=1$, TE $\ell=2$, and TM $\ell=3$ channels. (d) Fit of numerically calculated partial cross-sections into TM $\ell=1$, TE $\ell=2$, and TM $\ell=3$ scattering channel with CMT calculated cross-sections of a single-mode resonant system coupled to three scattering channels.

The two other modes correspond to transverse symmetric and anti-symmetric combinations of oscillating MDs and each is doubly degenerate with $m=\pm1$. They are formed by MDs oscillating in the particles in- and out-of-phase, respectively, orthogonal to the axis of the dimer. While the symmetric hybrid mode radiates mostly as a centered MD (which barely contributes to superscattering in the spectral range of interest, Fig. 2b), the anti-symmetric mode resembles a TD (in the near-field) and mainly couples to the ED (TM $\ell=1$) scattering channel; clearly, two anti-symmetric MDs also couple to the MQ channel. Thus, it is the anti-symmetric $m=\pm1$ mode of the dimer that should enable the superscattering regime via the multipole mixing [39].

To support this conclusion, we model light scattering by the dimer using coupled-mode theory (CMT) framework [55,56]. We describe the dimer as a single-mode linear system coupled to three scattering channels: TM $\ell=1$ (toroidal dipole), TE $\ell=2$ (magnetic quadrupole), and TM $\ell=3$ (electric octupole), Fig. 4c. The simplest system reproducing superscattering could be obtained by coupling the mode to two lowest channels, but we find that adding coupling to the electric octupole improves the agreement with numerical simulations. Fig. 4d shows partial scattering cross-sections into all three channels and the respective numerically calculated cross-section for $\ell=1$ and $\ell=2$ channels, all normalized by the dipole limit $\sigma_{max}^{\ell=1} = 3\lambda^2/(2\pi)$. Perfect agreement of CMT and full-wave simulations validates the multipole mixing mechanism of toroidal superscattering by the dielectric dimer.

Thus, the observed superscattering is, in fact, achieved via the combination of two mechanisms: 1) the conventional one due to overlapping resonances of different multipoles (i.e., TD and MQ) and 2) Friedrich-Wintgen mechanism of the super-multipole formation driving the violation of the single-channel scattering limit [39]. To the best of our knowledge, this is the first time when those two mechanisms are shown to work together.

## Experiment.

### 1.1 Materials& Fabrication

To demonstrate the effect of toroidal superscattering, we experimentally measured the total scattering cross-section in the gigahertz frequency range for a dimer composed of high-index ceramic cubes. The cubic shape of the particles was chosen to ease the handling of a brittle ceramic material during their fabrication. At the same time, the multipolar composition of the current mode excited in a cubic particle is nearly identical to that of a spherical particle, which allows replacing spheres with cubes without significant loss of accuracy of our model. The cubes were fabricated from commercially available dielectric ceramic Wangling TP-1/2 using a high-precision milling machine. The relative permittivity of the ceramic was 21 with the loss tangent tan δ ≤ 0.001 and the relative permittivity was 1. The cubes were embedded into Styrofoam material with air-like permittivity. The length of each side of the cubic particles was 8 mm and the distance between the centres of the cubes was 19 mm (see insert to Fig. 5b).

### 1.2 Setups for measuring near-field and extinction cross-section

All the measurements were performed in a microwave anechoic chamber. For near-field characterizations the sample was illuminated by a horn antenna HZ-137HA10NG. That antenna was connected to one of the ports of a vector network analyzer (VNA) Rohde & Schwarz ZVA-50 and used as a transmitting antenna. A receiving antenna was a custom-made loop probe, which was connected to the other port of the VNA. The probe was set to detect the z-component of the magnetic field in the near-field zone. (see Fig. 5a). The electromagnetic characteristics of the ceramic dimer were investigated in 6–7.5 GHz frequency range. The area covered with probe scanning was $70 \times 70$ mm$^2$ with a spatial resolution of 1mm.

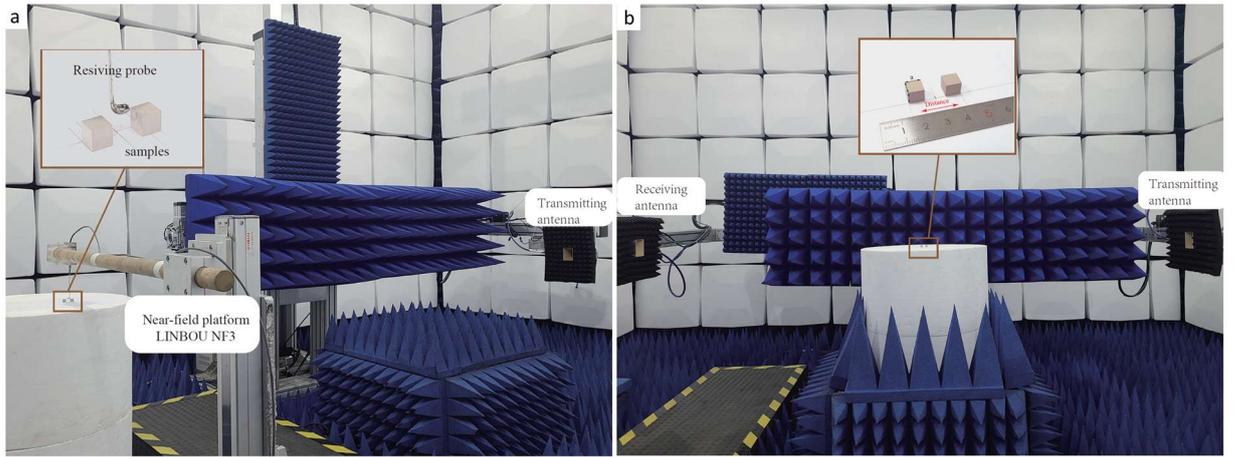

**Figure 5.** Photographs of the (a) near-field and (b) far-field (scattering cross-section) measurement setups. All measurements are carried out in an anechoic chamber covered by a radio frequency (RF) microwave absorber. In the inset of the scheme of the near-field setup (a), photographs of the magnetic dipole probe and sample of ceramically cubes are shown. The inset (b) shows samples of ceramic cubes with a scale bar showing the dimensions of the cubes and the distance between them.

A bistatic test bench was used to measure the extinction cross-section (see Fig. 5b). Two horn antennas were used as a transmitting and receiving antennas, respectively. They were connected to corresponding ports of the VNA through a low-noise amplifier (Ceyear-80223A) by 50 Ω coaxial cables. The center of the dimer was placed halfway between the antennas. The distance from the aperture of each antenna to the center of the dimer was 1.6 m. We used this setup to measure S-parameters of free space, a single cube and a dimer composed of two cubes. Free-space measurement yielded the background signal. The forward scattering amplitude was obtained as the difference between the transmission coefficient measured for two cubes and the background signal. Using the optical theorem, the extinction was extracted from the complex magnitude of the forward scattering signal [57]. The value of the scattering extinction was calibrated based on the value known for a single cube. The standard filtering procedure and time gating technique were used to reduce the noise and parasitic reverberations between the two antennas [32,58].

### 1.3 Experimental Validation of Toroidal Superscattering

To confirm the regime of toroidal superscattering, we compared the measured extinction cross-section and near-field distribution with the results of a full-wave simulation in the gigahertz frequency range. The numerical model was built using the exact dimensions and material properties of the ceramic dimer with the direction of the incident wave and its polarization being identical to those considered earlier in our optical model (as shown in Figure 1d).

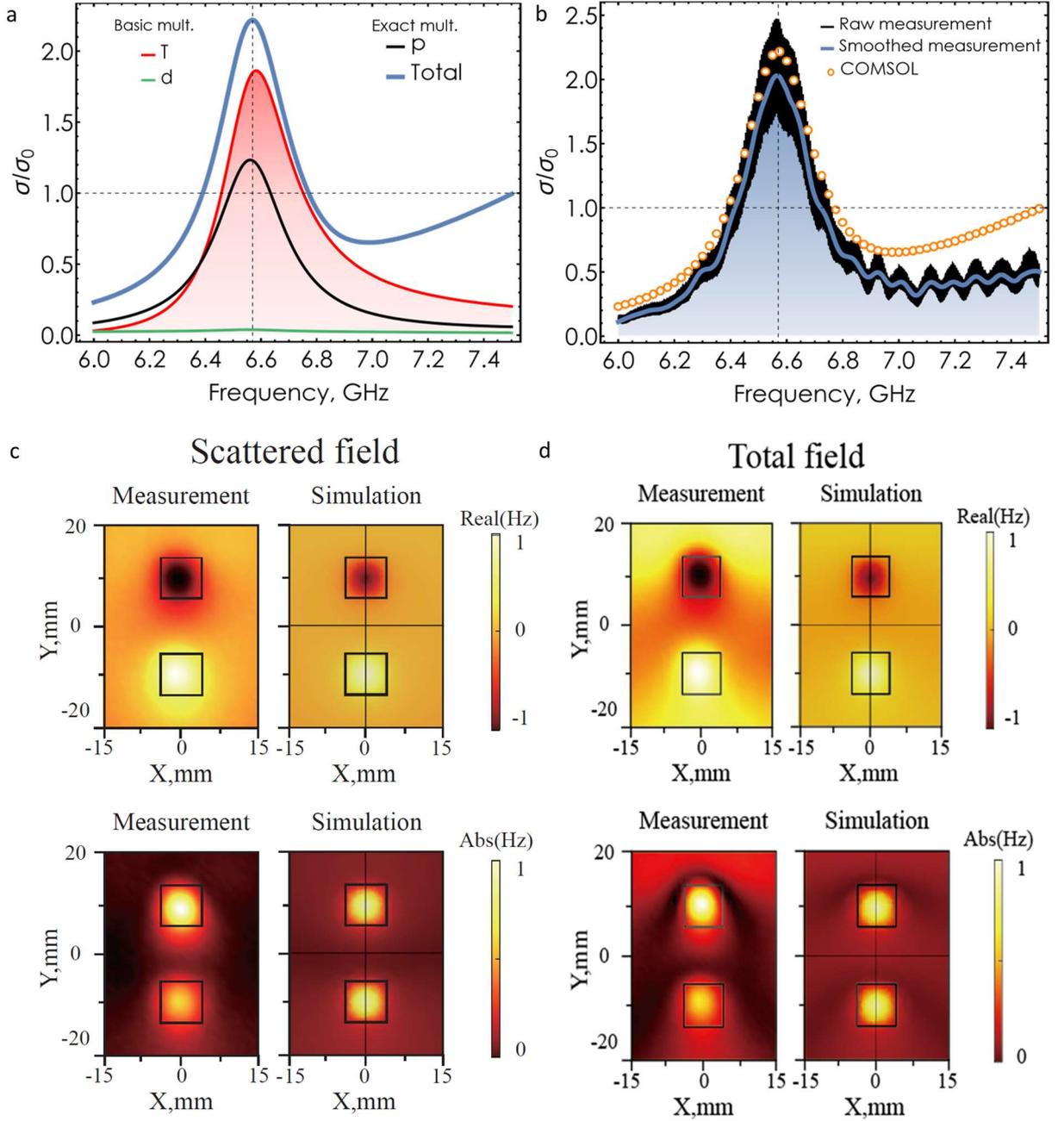

**Figure 6. a)** The multipole scattering cross-section for a dimer composed of high-index ceramic cubes is normalized to the limit of one dipole channel. p – "exact dipole", which includes toroidal components, d - *primitive ED*. T - scattering cross-section of a toroidal dipole only. All results obtained in COMSOL Multiphysics; b) measured (with and without smoothing) cross-section normalized by single-channel limit. Magnitude and real part of the magnetic field $H_z$ (c) scattered field and (d) total field. For comparison, the results of numerical simulations in COMSOL Multiphysics and direct measurements are presented. Frequency 6.575 GHz.

Figure 6a shows the normalized scattering cross-section obtained from COMSOL simulations, revealing a pronounced peak exceeding the single-channel dipole limit. Importantly, this peak corresponds to a dominant contribution of TD confirming our theoretical predictions. The experimentally measured frequency dependence of the scattering cross-section (Fig. 6b) reveals a distinct resonance at 6.575 GHz, which closely matches the frequency of TD superscattering obtained in the simulations. While minor discrepancies are observed (e.g., in the absolute peak value), they are attributable to fabrication tolerances and material losses. Furthermore, near-field mapping of the magnetic field ($H_z$ component) reveal a strong qualitative agreement between the simulation and experiment (Figs. 6c and 6d). Both exhibit characteristic features of anti-symmetric MDs yielding toroidal field topology in the near field. These observations serve as a compelling experimental evidence of the hybrid toroidal superscattering regime in a dimer composed of high-index dielectric cubes.

**Conclusion**

In this work, we demonstrated, both theoretically and experimentally, a novel mechanism of superscattering enabled by toroidal super-dipole excitation (far exceeding dipolar single-channel limit). Specifically, we designed and analyzed a simple dimer system composed of two high-index dielectric particles (spheres and cubes), which supports a strong toroidal dipole response under plane-wave illumination. The system was shown to exhibit a scattering cross-section significantly exceeding the aforementioned single-channel limit — a clear signature of superscattering.

We show for the first time that this enhancement arises from two complementary mechanisms: (1) constructive interference and coupling of quasi-normal modes within a single scattering channel — a realization of the Friedrich-Wintgen mechanism; and (2) spectral overlap of distinct multipoles, specifically toroidal dipole and magnetic quadrupole. Our coupled-mode theory and multipolar decomposition analyses clarify the interplay of these contributions, while full-wave simulations confirm their robustness. Importantly, we validated this phenomenon experimentally in the gigahertz regime using a dimer of ceramic cubes. The measured extinction spectra and near-field maps exhibit remarkable agreement with simulations and unambiguously confirm the predicted toroidal superscattering regime. Our results highlight that superscattering is not limited to electric or magnetic multipoles but can be realized via toroidal excitations being a separate family of multipoles, thus expanding the fundamental toolbox for manipulating light-matter interactions. The clear demonstration of toroidal super-scattering dipole fully coupled to an incident radiation could enable dynamic super-anapoles with novel peculiar optical signatures driven by both primitive and toroidal super-dipoles. This new phenomenon opens a venue for a variety of applications in toroidal electrodynamics, sensing, wireless communication, stealth technologies, and antenna engineering.

**Acknowledgements**

The calculations of non-Hermitian superscattering by a dimer of silicon spheres were partially supported by the grant from the Russian Science Foundation № 25-22-00266.

**The bibliography**

[1]     E. E. Radescu and G. Vaman, Exact calculation of the angular momentum loss, recoil force, and radiation intensity for an arbitrary source in terms of electric, magnetic, and toroid multipoles, Phys Rev E Stat Phys Plasmas Fluids Relat Interdiscip Topics **65**, (2002).

[2]     A. T. Góngora and E. Ley-Koo, Complete electromagnetic multipole expansion including toroidal moments, Revista Mexicana de Fisica E **52**, (2006).

[3]     V. Savinov, V. A. Fedotov, and N. I. Zheludev, Toroidal dipolar excitation and macroscopic electromagnetic properties of metamaterials, Phys Rev B Condens Matter Mater Phys **89**, (2014).

[4]     I. Fernandez-Corbaton, S. Nanz, and C. Rockstuhl, On the dynamic toroidal multipoles from localized electric current distributions, Sci Rep **7**, (2017).

[5]     G. N. Afanasiev and V. M. Dubovik, Electromagnetic properties of a toroidal solenoid, Journal of Physics A: General Physics **25**, (1992).

[6]     G. N. Afanasiev and V. M. Dubovik, Some remarkable charge-current configurations, Physics of Particles and Nuclei **29**, (1998).

[7]     M. Dubovik and A. Tosunyan, Axial toroidal moments in electrodynamics and solid-state physics, Journal of Experimental and Theoretical Physics **605**, (1986).

[8]     G. N. Afanasiev and Y. P. Stepanovsky, The electromagnetic field of elementary time-dependent toroidal sources, J Phys A Math Gen **28**, (1995).

[9]     V. A. Fedotov, A. V. Rogacheva, V. Savinov, D. P. Tsai, and N. I. Zheludev, Resonant transparency and non-trivial non-radiating excitations in toroidal metamaterials, Sci Rep **3**, (2013).

[10]    V. M. Dubovik, M. A. Martsenyuk, and B. Saha, Material equations for electromagnetism with toroidal polarizations, Phys Rev E Stat Phys Plasmas Fluids Relat Interdiscip Topics **61**, (2000).

[11]    A. E. Miroshnichenko, A. B. Evlyukhin, Y. F. Yu, R. M. Bakker, A. Chipouline, A. I. Kuznetsov, B. Luk'yanchuk, B. N. Chichkov, and Y. S. Kivshar, Nonradiating anapole modes in dielectric nanoparticles, Nat Commun **6**, 8069 (2015).

[12]    E. A. Marengo and R. W. Ziolkowski, Nonradiating sources, the Aharonov-Bohm effect, and the question of measurability of electromagnetic potentials, Radio Sci **37**, (2002).

[13]    A. Canós Valero et al., Theory, Observation, and Ultrafast Response of the Hybrid Anapole Regime in Light Scattering, Laser Photon Rev **15**, (2021).

[14]    A. V. Kuznetsov, A. C. Valero, M. Tarkhov, V. Bobrovs, D. Redka, and A. S. Shalin, Transparent hybrid anapole metasurfaces with negligible electromagnetic coupling for phase engineering, Nanophotonics **10**, (2021).

[15]    T. A. Raybould, V. A. Fedotov, N. Papasimakis, I. Kuprov, I. J. Youngs, W. T. Chen, D. P. Tsai, and N. I. Zheludev, Toroidal circular dichroism, Phys Rev B **94**, (2016).

[16]    J. S. Totero Gongora, A. E. Miroshnichenko, Y. S. Kivshar, and A. Fratalocchi, Anapole nanolasers for mode-locking and ultrafast pulse generation, Nat Commun **8**, (2017).

[17]    X. Zhang, J. Li, J. F. Donegan, and A. L. Bradley, Constructive and destructive interference of Kerker-type scattering in an ultrathin silicon Huygens metasurface, Phys Rev Mater **4**, (2020).

[18]    P. D. Terekhov, K. V. Baryshnikova, A. S. Shalin, A. Karabchevsky, and A. B. Evlyukhin, Resonant forward scattering of light by high-refractive-index dielectric nanoparticles with toroidal dipole contribution, Opt Lett **42**, (2017).

[19]    M. M. Bukharin, V. Y. Pecherkin, A. K. Ospanova, V. B. Il'in, L. M. Vasilyak, A. A. Basharin, and B. Luk'yanchuk, Transverse Kerker effect in all-dielectric spheroidal particles, Sci Rep **12**, (2022).

[20]    R. Peng, Q. Zhao, Y. Meng, and S. Wen, Pure toroidal dipole in a single dielectric disk, Opt Express **30**, (2022).

[21]    K. Marinov and V. A. Fedotov, Gyrotropy and permittivity sensing driven by toroidal response, New J Phys **25**, (2023).

[22]    C. K. Mididoddi, N. Papasimakis, and N. I. Zheludev, *The 18th International Congress on Artificial Materials for Novel Wave Phenomena – Metamaterials 2024*, in *Predator-Prey Nonreciprocal Interactions of Toroidal Charge-Current Configurations* (Crete, Greece, 2024).

[23]    N. A. Nemkov, A. A. Basharin, and V. A. Fedotov, Electromagnetic sources beyond common multipoles, Phys Rev A (Coll Park) **98**, (2018).

[24]    T. Kaelberer, V. A. Fedotov, N. Papasimakis, D. P. Tsai, and N. I. Zheludev, Toroidal dipolar response in a metamaterial, Science (1979) **330**, (2010).

[25]    Z. G. Dong, J. Zhu, J. Rho, J. Q. Li, C. Lu, X. Yin, and X. Zhang, Optical toroidal dipolar response by an asymmetric double-bar metamaterial, Appl Phys Lett **101**, (2012).

[26]    D. W. Watson, S. D. Jenkins, J. Ruostekoski, V. A. Fedotov, and N. I. Zheludev, Toroidal dipole excitations in metamolecules formed by interacting plasmonic nanorods, Phys Rev B **93**, (2016).

[27]    S. Guo, N. Talebi, A. Campos, M. Kociak, and P. A. Van Aken, Radiation of Dynamic Toroidal Moments, ACS Photonics **6**, (2019).

[28]    A. Canós Valero et al., On the Existence of Pure, Broadband Toroidal Sources in Electrodynamics, Laser Photon Rev **18**, (2024).

[29]    A. A. Basharin, M. Kafesaki, E. N. Economou, C. M. Soukoulis, V. A. Fedotov, V. Savinov, and N. I. Zheludev, Dielectric metamaterials with toroidal dipolar response, Phys Rev X **5**, (2015).

[30]    E. A. Gurvitz, K. S. Ladutenko, P. A. Dergachev, A. B. Evlyukhin, A. E. Miroshnichenko, and A. S. Shalin, The High-Order Toroidal Moments and Anapole States in All-Dielectric Photonics, Laser Photon Rev **13**, 1800266 (2019).

[31]    B. Luk'yanchuk, R. Paniagua-Domínguez, A. I. Kuznetsov, A. E. Miroshnichenko, and Y. S. Kivshar, Hybrid anapole modes of high-index dielectric nanoparticles, Phys Rev A (Coll Park) **95**, (2017).

[32] P. Kapitanova, E. Zanganeh, N. Pavlov, M. Song, P. Belov, A. Evlyukhin, and A. Miroshnichenko, Seeing the Unseen: Experimental Observation of Magnetic Anapole State Inside a High-Index Dielectric Particle, Ann Phys **532**, (2020).

[33] N. Papasimakis, V. A. Fedotov, V. Savinov, T. A. Raybould, and N. I. Zheludev, Electromagnetic toroidal excitations in matter and free space, Nat Mater **15**, (2016).

[34] Z. Ruan and S. Fan, Superscattering of light from subwavelength nanostructures, Phys Rev Lett **105**, (2010).

[35] Z. Ruan and S. Fan, Design of subwavelength superscattering nanospheres, Appl Phys Lett **98**, (2011).

[36] L. Verslegers, Z. Yu, Z. Ruan, P. B. Catrysse, and S. Fan, From electromagnetically induced transparency to superscattering with a single structure: A coupled-mode theory for doubly resonant structures, Phys Rev Lett **108**, (2012).

[37] V. I. Shcherbinin, V. I. Fesenko, T. I. Tkachova, and V. R. Tuz, Superscattering from Subwavelength Corrugated Cylinders, Phys Rev Appl **13**, (2020).

[38] C. Qian, X. Lin, Y. Yang, X. Xiong, H. Wang, E. Li, I. Kaminer, B. Zhang, and H. Chen, Experimental Observation of Superscattering, Phys Rev Lett **122**, (2019).

[39] A. Canós Valero, H. K. Shamkhi, A. S. Kupriianov, T. Weiss, A. A. Pavlov, D. Redka, V. Bobrovs, Y. Kivshar, and A. S. Shalin, Superscattering emerging from the physics of bound states in the continuum, Nat Commun **14**, (2023).

[40] A. V. Kuznetsov, A. Canós Valero, H. K. Shamkhi, P. Terekhov, X. Ni, V. Bobrovs, M. V. Rybin, and A. S. Shalin, Special scattering regimes for conical all-dielectric nanoparticles, Sci Rep **12**, (2022).

[41] H. Friedrich and D. Wintgen, Interfering resonances and bound states in the continuum, Phys Rev A  (Coll Park) **32**, (1985).

[42] Y. Yang, A. E. Miroshnichenko, S. V. Kostinski, M. Odit, P. Kapitanova, M. Qiu, and Y. S. Kivshar, Multimode directionality in all-dielectric metasurfaces, Phys Rev B **95**, (2017).

[43] K. Koshelev, S. Lepeshov, M. Liu, A. Bogdanov, and Y. Kivshar, Asymmetric Metasurfaces with High- Q Resonances Governed by Bound States in the Continuum, Phys Rev Lett **121**, (2018).

[44] Z. F. Sadrieva, I. S. Sinev, K. L. Koshelev, A. Samusev, I. V. Iorsh, O. Takayama, R. Malureanu, A. A. Bogdanov, and A. V. Lavrinenko, Transition from Optical Bound States in the Continuum to Leaky Resonances: Role of Substrate and Roughness, ACS Photonics **4**, (2017).

[45] A. B. Evlyukhin, C. Reinhardt, A. Seidel, B. S. Luk'Yanchuk, and B. N. Chichkov, Optical response features of Si-nanoparticle arrays, Phys Rev B Condens Matter Mater Phys **82**, (2010).

[46] O. Merchiers, F. Moreno, F. González, and J. M. Saiz, Light scattering by an ensemble of interacting dipolar particles with both electric and magnetic polarizabilities, Phys Rev A **76**, (2007).

[47] A. K. Ospanova, A. Basharin, A. E. Miroshnichenko, and B. Luk'yanchuk, Generalized hybrid anapole modes in all-dielectric ellipsoid particles [Invited], Opt Mater Express **11**, (2021).

[48] K. Ladutenko, P. Belov, O. Peña-Rodríguez, A. Mirzaei, A. E. Miroshnichenko, and I. V. Shadrivov, Superabsorption of light by nanoparticles, Nanoscale **7**, (2015).

[49] D. C. Marinica, A. G. Borisov, and S. V. Shabanov, Bound states in the continuum in photonics, Phys Rev Lett **100**, (2008).

[50] C. W. Hsu, B. Zhen, J. Lee, S. L. Chua, S. G. Johnson, J. D. Joannopoulos, and M. Soljačić, Observation of trapped light within the radiation continuum, Nature **499**, (2013).

[51] A. A. Bogdanov, K. L. Koshelev, P. V. Kapitanova, M. V. Rybin, S. A. Gladyshev, Z. F. Sadrieva, K. B. Samusev, Y. S. Kivshar, and M. F. Limonov, Bound states in the continuum and Fano resonances in the strong mode coupling regime, Advanced Photonics **1**, (2019).

[52] M. V. Rybin, K. L. Koshelev, Z. F. Sadrieva, K. B. Samusev, A. A. Bogdanov, M. F. Limonov, and Y. S. Kivshar, High- Q Supercavity Modes in Subwavelength Dielectric Resonators, Phys Rev Lett **119**, (2017).

[53] C. , Z. B. , S. A. et al. Hsu, Bound states in the continuum, Nat Rev Mater **1**, 16048 (2016).

[54] S. Krasikov et al., Multipolar Engineering of Subwavelength Dielectric Particles for Scattering Enhancement, Phys Rev Appl **15**, (2021).

[55] R. E. Hamam, A. Karalis, J. D. Joannopoulos, and M. Soljačić, Coupled-mode theory for general free-space resonant scattering of waves, Phys Rev A **75**, (2007).

[56] Z. Ruan and S. Fan, Temporal coupled-mode theory for light scattering by an arbitrarily shaped object supporting a single resonance, Phys Rev A **85**, (2012).

[57] R. G. Newton, Optical theorem and beyond, Am J Phys **44**, (1976).

[58] M. Odit, P. Kapitanova, P. Belov, R. Alaee, C. Rockstuhl, and Y. S. Kivshar, Experimental realisation of all-dielectric bianisotropic metasurfaces, Appl Phys Lett **108**, (2016).

# Table of content

## Hybrid Superscattering Driven by Toroidal Dipole

Kislov[1,2,3,*] D., Borovkov[1] D., Huang[3] L., Kuznetsov[1] A., Canos Valero[4] A., Ipatovs[5] A., Bobrovs[5] V., Fedotov[2,*] V., Gao[6] L., Xie[3] S., Xu[7] Y., Luo[8] J., Baranov[1] D., Arsenin[1] A., Bolshakov[1] A., Shalin[1,2,3,6,*] A.S.

[1]Center for Photonics and 2D Materials, Moscow Institute of Physics and Technology, Dolgoprudny 141700, Russia
[2]Centre for Photonic Science and Engineering, Skolkovo Institute of Science and Technology, Moscow, 121125, Russia.
[3]Beijing Institute of Technology Beijing 100081, China
[4]Institute of Physics, University of Graz, and NAWI Graz, 8010, Graz, Austria
[5] Riga Technical University, Institute of Photonics, Electronics and Telecommunications, 1048 Riga, Latvia
[6]School of Optical and Electronic Information, Suzhou City University, Suzhou 215104, China.
[7]Guangdong University of Technology, Guangzhou 510006, China
[8] School of Physical Science and Technology, Soochow University, Suzhou 215006, China

*Corresponding Author E-Mail: alexandesh@gmail.com; denis.a.kislov@gmail.com; V.Fedotov@skoltech.ru

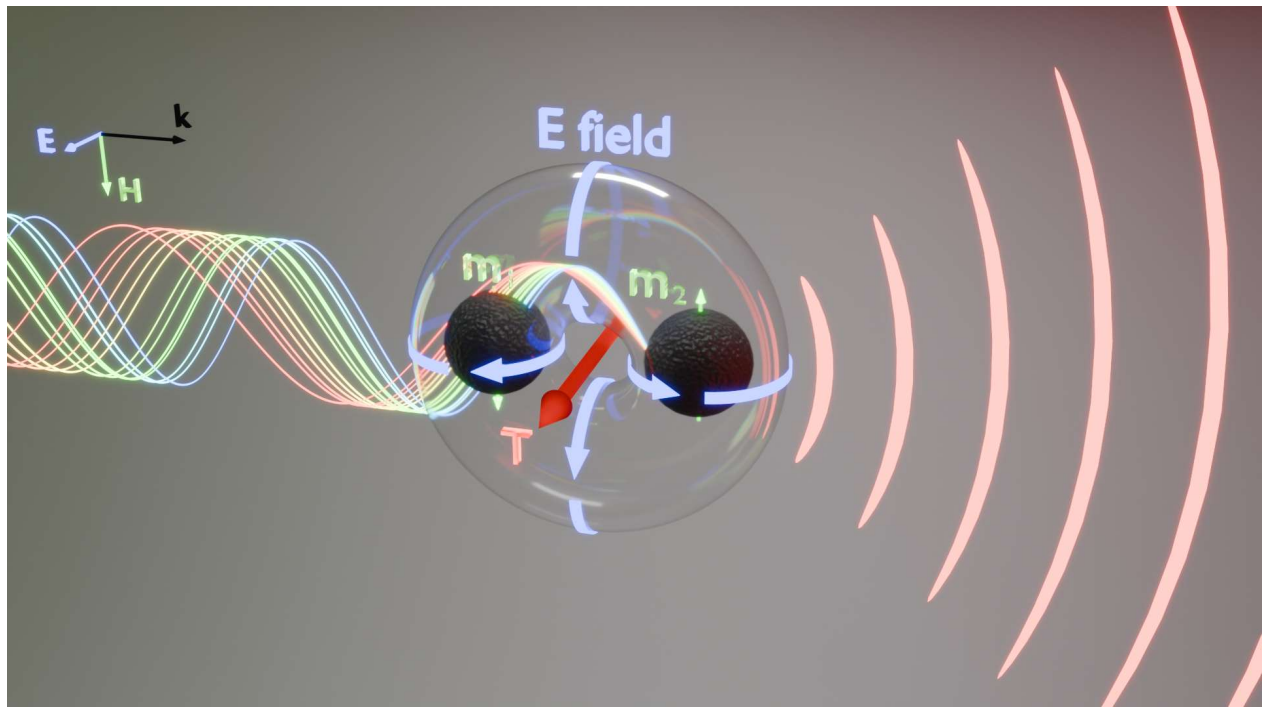

Illustration of a dimer formed by two Si spheres supporting transverse anti-symmetric combinations of oscillating MDs, respectively, orthogonal to the axis of the dimer. The antisymmetric mode of the dimer provides a superscattering regime via multipole mixing of the toroidal electric dipole and magnetic quadrupole.

# Supplementary Information

# Hybrid Superscattering Driven by Toroidal Dipole

Kislov[1,2,3,*] D., Borovkov[1] D., Huang[1] L., Kuznetsov[1] A., Canos Valero[4] A., Ipatovs[5] A., Bobrovs[5] V., Fedotov[2,*] V., Gao[6] L., Xie[3] S., Xu[7] Y., Luo[8] J., Baranov[1] D., Arsenin[1] A., Bolshakov[1,2,3,6,*] A.S.

[1]Center for Photonics and 2D Materials, Moscow Institute of Physics and Technology, Dolgoprudny 141700, Russia
[2]Centre for Photonic Science and Engineering, Skolkovo Institute of Science and Technology, Moscow, 121125, Russia.
[3]Beijing Institute of Technology Beijing 100081, China
[4]Institute of Physics, University of Graz, and NAWI Graz, 8010, Graz, Austria
[5] Riga Technical University, Institute of Photonics, Electronics and Telecommunications, 1048 Riga, Latvia
[6]School of Optical and Electronic Information, Suzhou City University, Suzhou 215104, China.
[7]Guangdong University of Technology, Guangzhou 510006, China
[8] School of Physical Science and Technology, Soochow University, Suzhou 215006, China

*Corresponding Author E-Mail: alexandesh@gmail.com; denis.a.kislov@gmail.com; V.Fedotov@skoltech.ru

## S1. Coupled-dipole approximation

The particles are considered as electric and magnetic dipoles

$$\mathbf{p} = \alpha^E \mathbf{E}^0, \quad \mathbf{m} = \alpha^M \mathbf{H}^0, \tag{S1.1}$$

with polarizabilities $\alpha^E$ and $\alpha^M$ respectively

$$\alpha^E = i\frac{6\pi\varepsilon_0\varepsilon_d}{k_d^3}a_1, \quad \alpha^M = i\frac{6\pi}{k_d^3}b_1, \tag{S1.2}$$

the dimensionless Mie coefficients $a_1$ and $b_1$ are expressed by using the Riccati-Bessel functions.

The particles are placed in the dielectric medium with electric permittivity $\varepsilon_d$ and illuminated with a plane wave with electric $\mathbf{E}^0(\mathbf{r})\exp(-i\omega t) = \mathbf{E}^0\exp(i\mathbf{k}_d\mathbf{r} - i\omega t)$ and magnetic $\mathbf{H}^0(\mathbf{r})\exp(-i\omega t) = \mathbf{H}^0\exp(i\mathbf{k}_d\mathbf{r} - i\omega t)$ fields (the wave vector $\mathbf{k}_d$ is given in the embedding dielectric medium with $\varepsilon_d$ and $\omega$ is circular frequency of the wave). For two particles the system of equations is reduced to

$$\mathbf{p}_1 = \alpha^E\left[\mathbf{E}_1^0 + \frac{k_0^2}{\varepsilon_0}\left(\hat{G}\mathbf{p}_2 - \frac{i}{ck_0}[\mathbf{g}\times\mathbf{m}_2]\right)\right],$$
$$\mathbf{p}_2 = \alpha^E\left[\mathbf{E}_2^0 + \frac{k_0^2}{\varepsilon_0}\left(\hat{G}\mathbf{p}_1 + \frac{i}{ck_0}[\mathbf{g}\times\mathbf{m}_1]\right)\right], \tag{S1.3}$$

for the electric dipoles of particles and

$$\mathbf{m}_1 = \alpha^M\left[\mathbf{H}_1^0 + k_0^2\left(\varepsilon_d\hat{G}\mathbf{m}_2 + \frac{ic}{k_0}[\mathbf{g}\times\mathbf{p}_2]\right)\right],$$
$$\mathbf{m}_2 = \alpha^M\left[\mathbf{H}_2^0 + k_0^2\left(\varepsilon_d\hat{G}\mathbf{m}_1 - \frac{ic}{k_0}[\mathbf{g}\times\mathbf{p}_1]\right)\right], \tag{S1.4}$$

for the magnetic ones. $\mathbf{E}_j^0 = \mathbf{E}^0(\mathbf{r}_j)$ and $\mathbf{H}_j^0 = \mathbf{H}^0(\mathbf{r}_j)$ are incident electric and magnetic fields at the point of the particle $j$, $k_0$ is the wavenumber in vacuum, $c = \sqrt{\varepsilon_0 \mu_0}$ is the speed of light with $\varepsilon_0$ and $\mu_0$ are the vacuum permittivity and permeability, respectively. $\hat{G}$ is the Green's tensor of the medium without particles

$$\hat{G} = \left[ \left( \frac{1}{R} + \frac{i}{k_d R^2} - \frac{1}{k_d^2 R^3} \right) \hat{U} + \left( -\frac{1}{R} - \frac{3i}{k_d R^2} + \frac{3}{k_d^2 R^3} \right) \mathbf{e_R} \otimes \mathbf{e_R} \right] \frac{e^{ik_d R}}{4\pi}. \tag{S1.5}$$

Here $R = |\mathbf{R}_{12}| = |\mathbf{r}_1 - \mathbf{r}_2|$ is the distance between particles, $\mathbf{e_R} \otimes \mathbf{e_R}$ is the dyadic formed with the unit vector $\mathbf{e_R} = \mathbf{R}_{12} / R$ and $\hat{U}$ is the unit $3 \times 3$ tensor. And the vector $\mathbf{g}$ has the expression

$$\mathbf{g} = \mathbf{g}_{12} = -\mathbf{g}_{21} = \frac{e^{ik_d R}}{4\pi} \left( \frac{ik_d}{R} - \frac{1}{R^2} \right) \mathbf{e_R}. \tag{S1.6}$$

## S2. The total normalized cross-section of dimer for different distances between spheres

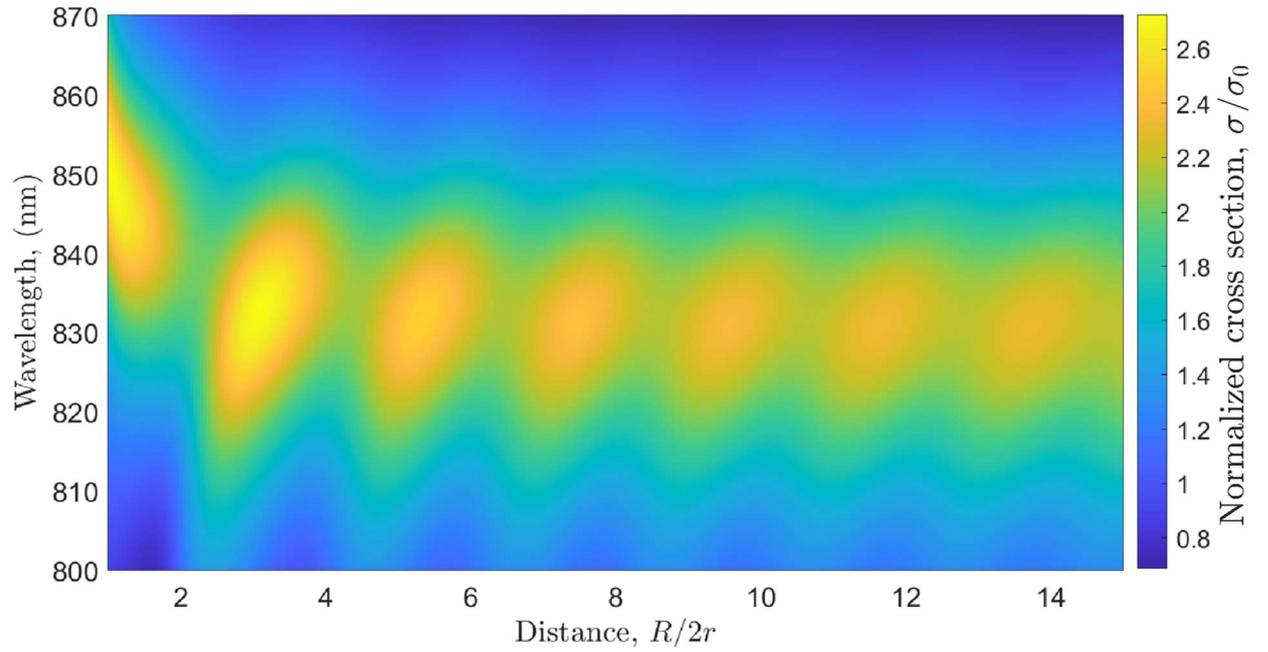

**Figure S2.1.** a) The total normalized cross-section of the system of two spheres mapped for different (normalised) distances between them and wavelengths in the vicinity of MD resonance of a single sphere. At distances $R/2r > 2$ the total scattering cross-section oscillates due to weak interaction between the particles. However, at distances $R/2r < 2$ the interaction kicks in and the scattering peak is seen to red-shift.

**S3. Multipolar decomposition.**

Let us write down the first few orders of exact multipoles expressed in Cartesian coordinates [1]. These expressions were used in the scattering modeling in the COMSOL Multiphysics package.

**Table S3.1. The first three orders of exact multipoles in Cartesian coordinates**

| Multipole type | Expression |
|---|---|
| Electric dipole | $p_\alpha = \dfrac{i}{\omega}\left\{ \int J_\alpha^\omega j_0(kr)d^3\mathbf{r} + \dfrac{k^2}{2}\iint \left[ 3(\mathbf{r}\cdot\mathbf{J}^\omega)r_\alpha - r^2\mathbf{J}_\alpha^\omega \right]\dfrac{j_2(kr)}{(kr)^2}d^3\mathbf{r} \right\}$ (S3.1) |
| Magnetic dipole | $m_\alpha = \dfrac{3}{2}\int [\mathbf{r}\times\mathbf{J}^\omega]_\alpha \dfrac{j_1(kr)}{kr}d^3\mathbf{r}$ $\qquad$ (S3.2) |
| Electric quadrupole | $Q_{\alpha\beta}^e = \dfrac{3i}{\omega}\left\{ \int \left[ 3(r_\beta J_\alpha^\omega + r_\alpha J_\beta^\omega) - 2\left( \mathbf{r}\cdot\mathbf{J}^\omega \right)\delta_{\alpha\beta} \right]\dfrac{j_1(kr)}{kr}d^3\mathbf{r} + \right.$ $\left. +2k^2\int \left[ 5r_\alpha r_\beta \left( \mathbf{r}\cdot\mathbf{J}^\omega \right) - (r_\beta J_\alpha^\omega + r_\alpha J_\beta^\omega)r^2 - r^2\left( \mathbf{r}\cdot\mathbf{J}^\omega \right)\delta_{\alpha\beta} \right]\dfrac{j_3(kr)}{(kr)^3}d^3\mathbf{r} \right\}$ (S3.3) |
| Magnetic quadrupole | $Q_{\alpha\beta}^m = 15\int \left\{ r_\alpha \left[ \mathbf{r}\times\mathbf{J}^\omega \right]_\beta + r_\beta \left[ \mathbf{r}\times\mathbf{J}^\omega \right]_\alpha \right\}\dfrac{j_2(kr)}{(kr)^2}d^3\mathbf{r}$ $\quad$ (S3.4) |

where $\alpha,\beta,\gamma \in \{x,y,z\}$; $\mathbf{J}^\omega = \mathbf{J}^\omega(\mathbf{r})$- is the spatially localized electric current density distribution; ω - light angular frequency; $k$ - is the wavenumber of light; $j_n(kr)$ - is a spherical Bessel function of the n-th order.

To reveal the nature of toroidal superscattering, we used expressions for the basic multipoles and their toroidal corrections of different orders [2,3].

**Table S3.2. Basic electric dipole and its higher order toroidal corrections**

| Multipole type | Expression |
|---|---|
| Electric dipole | $d_\alpha = \dfrac{i}{\omega}\int J_\alpha^\omega d^3\mathbf{r}$ <br> (S3.5) |
| Toroidal dipole | $T_\alpha = \dfrac{1}{10}\int \left[ (\mathbf{J}^\omega \cdot \mathbf{r})r_\alpha - 2r^2 J_\alpha^\omega \right]d^3\mathbf{r}$ $\qquad$ (S3.6) |
| Mean-square radius | $MSR = \dfrac{1}{280}\int \left[ 3r^4 J_\alpha^\omega - 2r^2(\mathbf{J}^\omega \cdot \mathbf{r})J_\alpha^\omega \right]d^3\mathbf{r}$ $\quad$ (S3.7) |

The total scattering cross-section of nanoparticle in terms of exact multipole moments up to the quadrupoles is given by:

$$\sigma_{sca} = \frac{k_0^2}{\pi \varepsilon_0^2 \left| \mathbf{E}_{inc} \right|^2} \left[ \frac{1}{6} \sum_{\alpha=x,y,z} \left( \left| p_\alpha \right|^2 + \left| \frac{m_\alpha}{c} \right|^2 \right) + \frac{1}{720} \sum_{\alpha,\beta=x,y,z} \left( \left| k_0 Q^e_{\alpha\beta} \right|^2 + \left| \frac{k_0 Q^m_{\alpha\beta}}{c} \right|^2 \right) + \dots \right]$$

$$(S3.8)$$

where $\left| \mathbf{E}_{inc} \right|$ - is the electric field amplitude of the incident wave., $k_0$ - is the wavenumber in vacuum, $\varepsilon_0$ is the vacuum permittivity and $c$ is the speed of light.

The dipole scattering cross-section can be described in terms of basic multipoles and toroidal terms [3]:

$$\sigma^p_{sca} \approx \frac{k^2}{6\pi \varepsilon_0^2 \left| \mathbf{E}_{inc} \right|^2} \sum_{\alpha=x,y,z} \left( \left| d_\alpha + \frac{ik}{c} T_\alpha + \frac{ik^3}{c} MSR_\alpha \right|^2 \right) \qquad (S3.9)$$

**S.4. Derivation of the higher modes conditions for a set of point magnetic dipoles (PMD).**

The current for each PMD is defined as the magnetization rotor multiplied by the δ-function. Thus, the total current of this system is:

$$\mathbf{J} = \sum_j \mathbf{J}^j = \sum_j \nabla \times \left( \mathbf{m}^j \delta(\mathbf{r} - \mathbf{r}^j) \right) \tag{S4.1}$$

$\mathbf{r}^j$ - position of a point magnetic dipole. In component notation, this expression has the form:

$$J_\alpha = \sum_j \varepsilon_{\alpha\beta\gamma} \partial_\beta (m_\gamma^j \delta(\mathbf{r} - \mathbf{r}^j)) \tag{S4.2}$$

To calculate an TD-like moment for the PMD with number $j$, we will put this expression into the integral in Eq.(S3.6):

$$T_a = -\frac{1}{i\omega} \frac{k^2}{10} \int dV \left[ (\mathbf{r} \cdot \mathbf{J}_\omega) r_\alpha - 2r^2 J_\alpha^\omega \right] = -\frac{1}{i\omega} \frac{k^2}{10} \int dV \left[ (r_\mu J_\mu^\omega) r_\alpha - 2r_\mu r_\mu J_\alpha^\omega \right] =$$

$$= -\frac{1}{i\omega} \frac{k^2}{10} \left[ \sum_j \varepsilon_{\mu\beta\gamma} m_\gamma^j \int r_\mu r_\alpha \partial_\beta (\delta(\mathbf{r} - \mathbf{r}^j)) dV - \sum_j 2\varepsilon_{\alpha\beta\gamma} m_\gamma^j \int r_\mu r_\mu \partial_\beta (\delta(\mathbf{r} - \mathbf{r}^j)) dV \right]$$

(S4.3)

Here and below we imply the use of Einstein's rule - summation over repeating indices. Next, using the definition of the derivative of the δ-function $\int_{-\infty}^{+\infty} f(x)\delta'(x-a)dx = -f'(a)$, the properties of the Kronecker symbol and performing some algebraic manipulations, we get rid of the integrals in the expression:

$$T_a = -\frac{1}{i\omega} \frac{k^2}{10} \left[ \sum_j \left( -\varepsilon_{\mu\beta\gamma} m_\gamma^j r_\alpha^j \delta_{\mu\beta} - \varepsilon_{\mu\beta\gamma} m_\gamma^j r_\mu^j \delta_{\alpha\beta} \right) - \sum_j \left( -2\varepsilon_{\alpha\beta\gamma} m_\gamma^j r_\mu^j \delta_{\mu\beta} - 2\varepsilon_{\alpha\beta\gamma} m_\gamma^j r_\mu^j \delta_{\mu\beta} \right) \right]$$

(S4.4)

Then, by folding the Levi-Civita symbol and the Kronecker symbol by the repeating index and rearranging the terms, we obtain the final expression:

$$T_a = -\frac{1}{i\omega} \frac{k^2}{2} \left[ \sum_j \left( \varepsilon_{\alpha\mu\gamma} m_\gamma^j r_\mu^j \right) \right] \tag{S4.5}$$

It can be noted that the expression (S4.5) is a component-wise notation of the following expression:

$$\mathbf{T} = -\frac{1}{i\omega} \frac{k^2}{2} \left[ \sum_j \left[ \mathbf{r}^j \times \mathbf{m}^j \right] \right] \tag{S4.6}$$

Using a similar approach, we can obtain an expression for the magnetic quadrupole moment expressed in terms of PMDs. To do this, we substitute the expression for the total current (S4.2) into the expression for the magnetic quadrupole moment:

$$Q_\alpha^m = \int d^3 \mathbf{r} \left\{ r_\alpha (\mathbf{r} \times \mathbf{J}_\omega)_\beta + r_\beta (\mathbf{r} \times \mathbf{J}_\omega)_\alpha \right\} = \int d^3 \mathbf{r} \left\{ r_\alpha \varepsilon_{\beta ij} r_i J_j^\omega + r_\beta \varepsilon_{\alpha kl} r_k J_l^\omega \right\} =$$

$$= \sum_s \varepsilon_{\beta ij} \varepsilon_{jmn} m_n^s \int dV \left\{ r_\alpha r_i \partial_m (\delta(\mathbf{r} - \mathbf{r}^s)) \right\} + \sum_s \varepsilon_{\alpha kl} \varepsilon_{lpq} m_q^s \int dV \left\{ r_\beta r_k \partial_p (\delta(\mathbf{r} - \mathbf{r}^s)) \right\}$$

$$\text{(S4.7)}$$

Next, performing a series of algebraic manipulations, we get rid of the integrals and, having regrouped the terms, we obtain the final expression:

$$Q_{\alpha\beta}^m = \sum_s 3 \left( r_\alpha^s m_\beta^s + r_\beta^s m_\alpha^s \right) - 2 \left( \mathbf{m}^s \cdot \mathbf{r}^s \right) \delta_{\alpha\beta} \tag{S4.8}$$

Let's check the obtained expression for tracelessness. To do this, we calculate the trace of the obtained tensor:

$$\mathbf{Tr} \left\{ Q_{\alpha\beta}^m \right\} = 3 \left( r_x^s m_x^s + r_x^s m_x^s \right) - 2 \left( r_x^s m_x^s + r_y^s m_y^s + r_z^s m_z^s \right) + 3 \left( r_y^s m_y^s + r_y^s m_y^s \right) - 2 \left( r_x^s m_x^s + r_y^s m_y^s + r_z^s m_z^s \right) +$$

$$+ 3 \left( r_z^s m_z^s + r_z^s m_z^s \right) - 2 \left( r_x^s m_x^s + r_y^s m_y^s + r_z^s m_z^s \right) =$$

$$= 6 r_x^s m_x^s + 6 r_y^s m_y^s + 6 r_z^s m_z^s - 6 \left( r_x^s m_x^s + r_y^s m_y^s + r_z^s m_z^s \right) = 0$$

$$\text{(S4.9)}$$

As expected, the magnetic quadrupole tensor turned out to be traceless.

The results obtained within the framework of this analytical model show that the determining contribution to the dimer scattering will be given by the toroidal dipole and magnetic quadrupole. As can be seen from Figure S4.1, the agreement between the model and the exact numerical simulation is good.

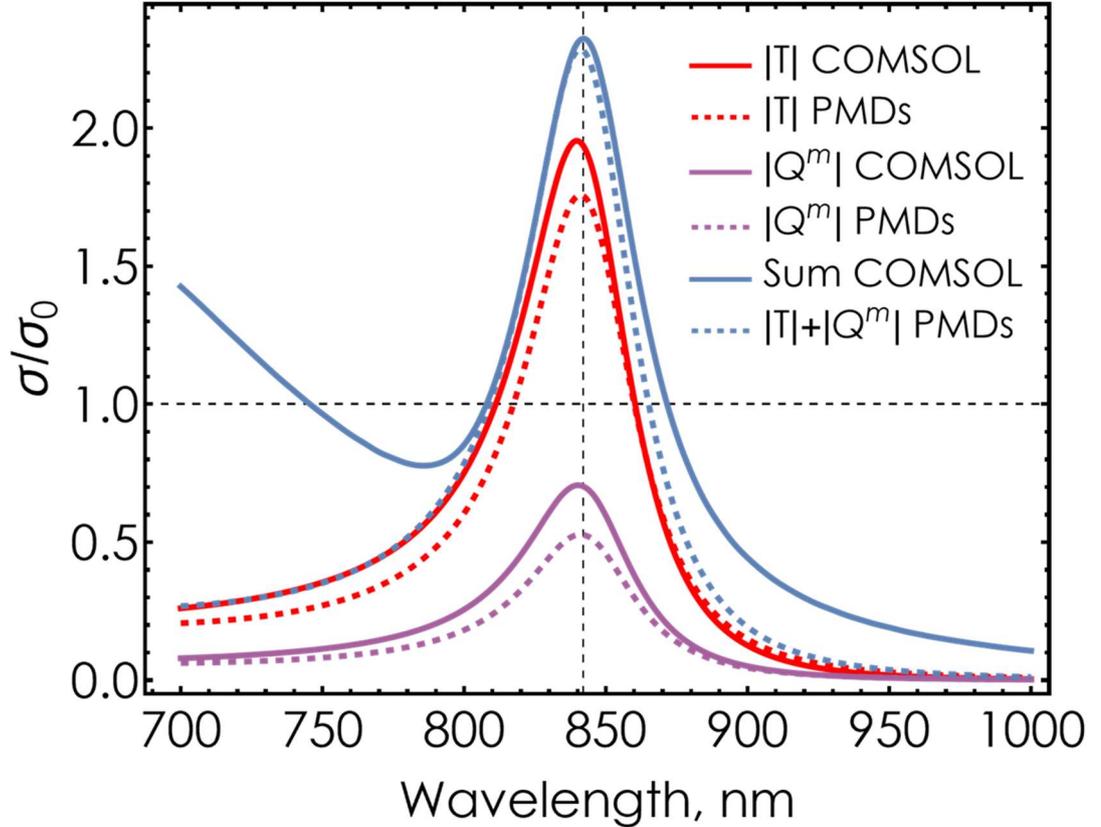

**Figure S4.1.** The normalized scattering cross-section of TD and MQ, of a dimer. Solid lines are exact numerical simulations, dotted lines are analytical calculations based on the PDMs model.

**S5. Mode analyse single spherical nanoscatterer**

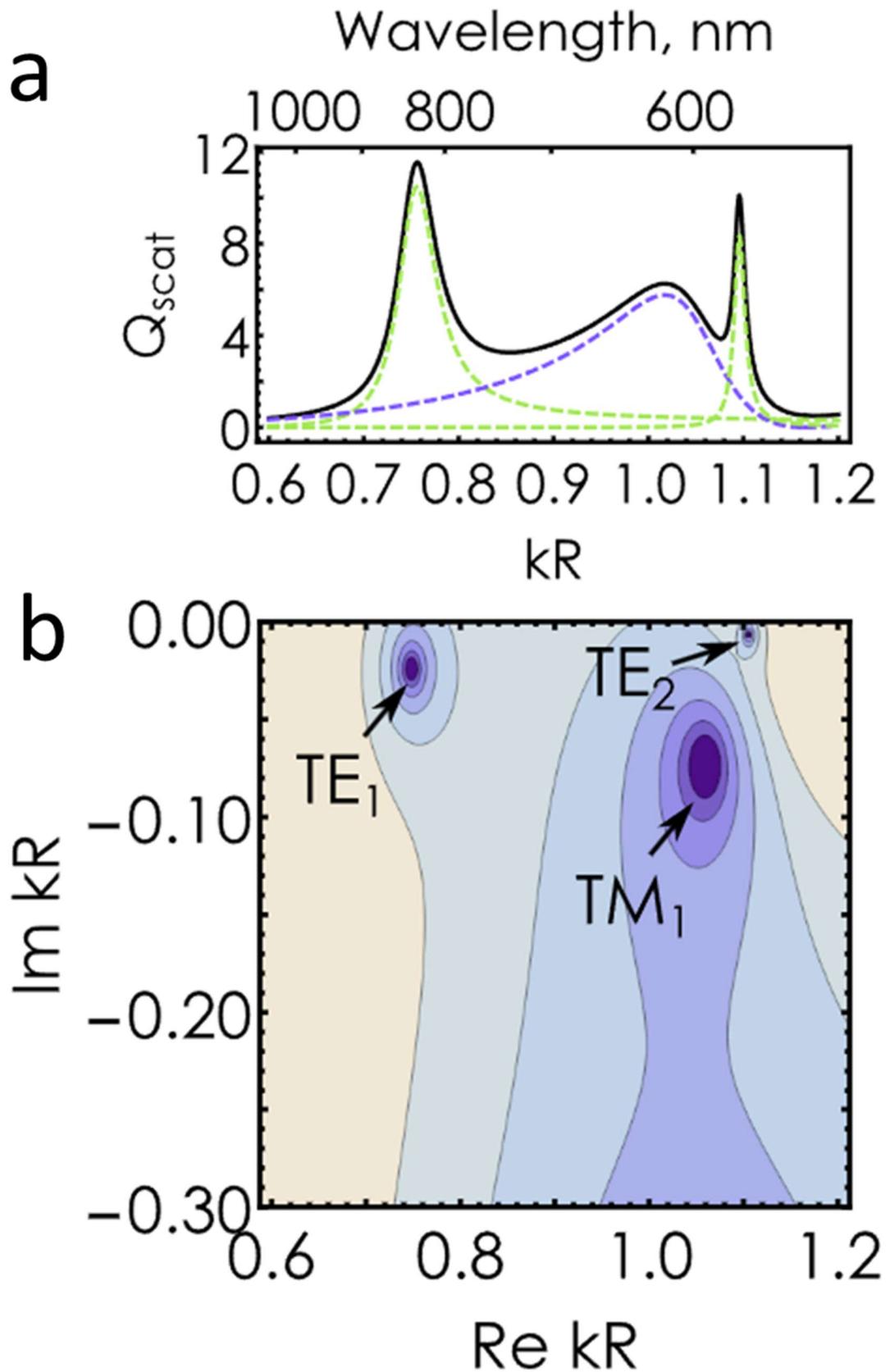

**Figure S5.1.** (a) Scattering spectrum multipole decomposition and (b) complex-valued eigenfrequencies of Si nanosphere with radius of 100 nm and refractive index *n* = 4.

## S6. Mode analysis of dimer.

The strong interaction between the MD particles in the dimer gives rise to four hybrid modes in the spectral region under study, with the eigenfrequencies exhibiting nontrivial dependences on the change distance between the spheres (see Fig. S6.1).

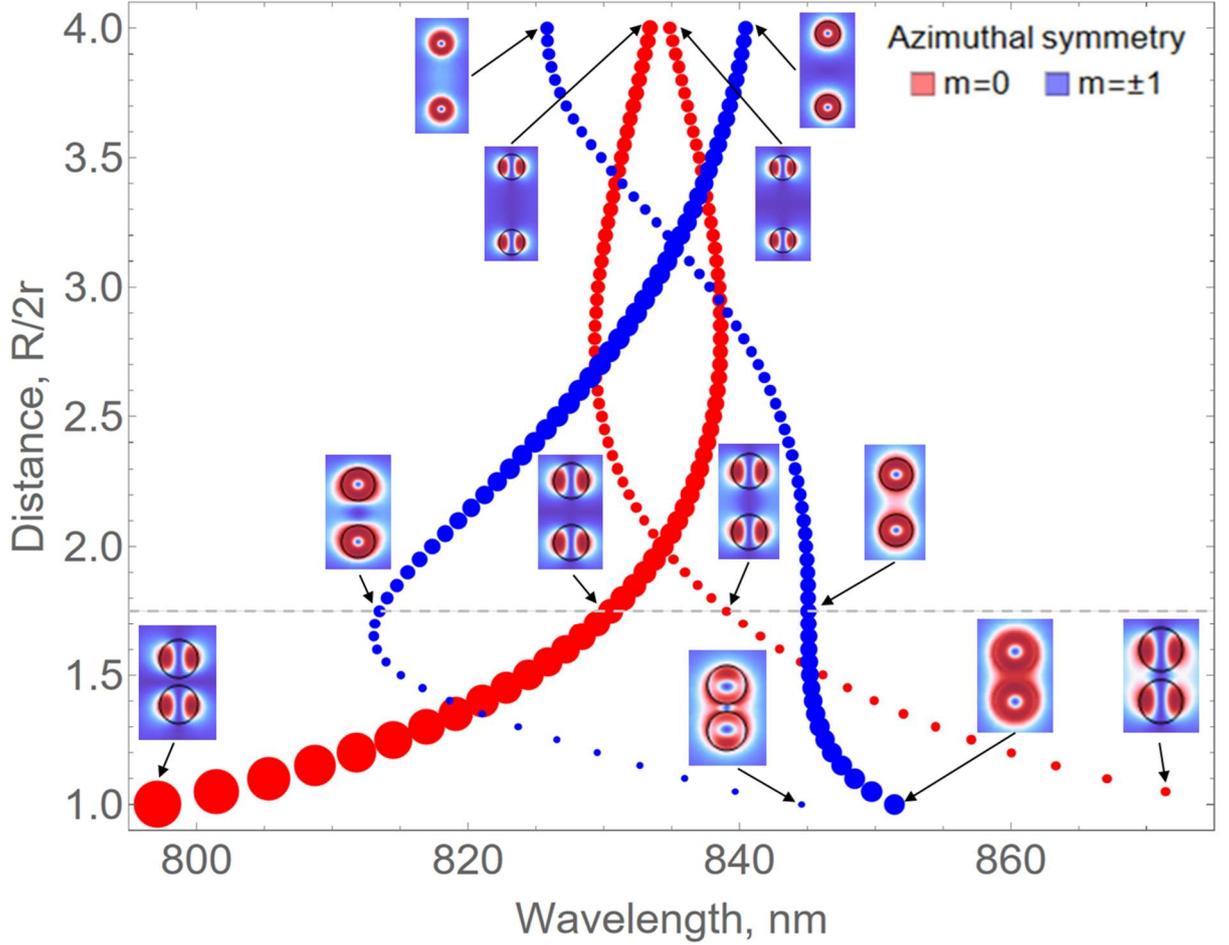

**Figure S6.1** Complex-valued eigenfrequencies of a dimer of two dielectric $n=4$ particles of radius $r$ at a center-to-center distance $R$ in the spectral range of interest close to the fundamental MD resonance of the single particle. The size of the dots corresponds to the Q- factor of the mods. The insets show the spatial distribution of the electric field norm for all four eigenmodes. The field distributions for several values of the distances between the spheres are shown.

In particular, two eigenmodes have azimuthal symmetry ($m = 0$). These are longitudinal anti-symmetric and symmetric hybrid modes formed by MDs oscillating in the particles in- and out-of-phase, respectively, along the axis of the dimer (see the distribution of H fields Fig. S6.2). In our case such modes appear as dark since they cannot be excited by a plane wave propagating along the axis of the dimer. In fact, due to its radial symmetry, the symmetric mode cannot be coupled to with a plane wave regardless of the propagation direction. For these reasons we have excluded the longitudinal modes from our further analysis.

The two other modes are characterised by $m = \pm 1$ and correspond to transverse symmetric and anti-symmetric hybrid modes. They are formed by MDs oscillating in the particles in- and out-of-phase, respectively, orthogonal to the axis of the dimer.

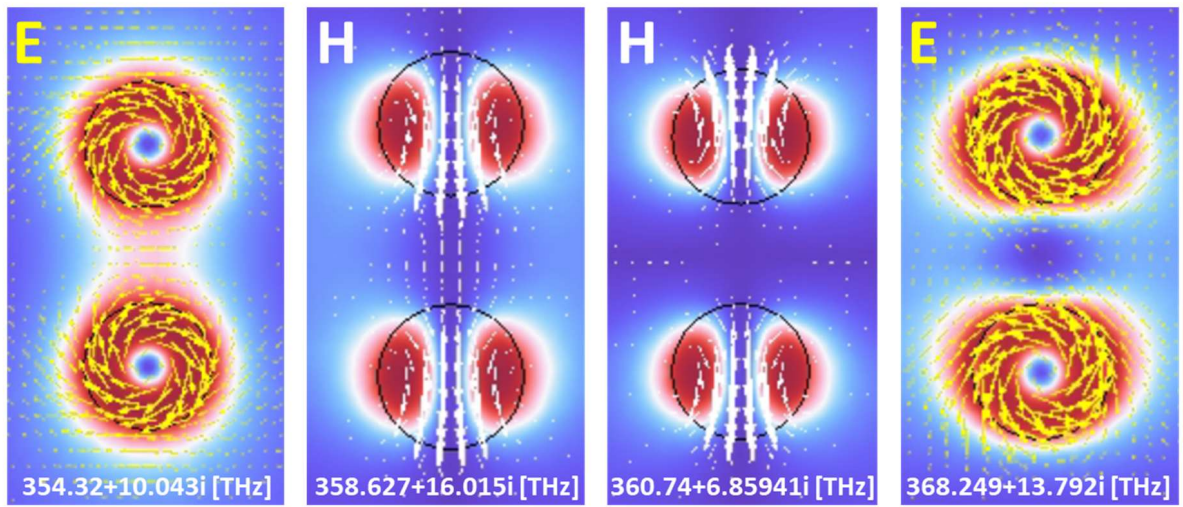

**Figure S6.2** Distribution of the electric field norm of four modes. For convenience of determining symmetric and anti-symmetric hybrid modes, the arrows show the direction of the electric and magnetic fields. Yellow arrows - electric field (for modes with azimuthal symmetry m=±1), white arrows - magnetic field (for modes with azimuthal symmetry m=0).

The vector distribution of the electric or magnetic fields enables determination of whether the induced magnetic dipoles in the nanospheres oscillate in phase or out of phase.

## S7. Temporal-coupled mode theory (TCMT)

Resonant scattering by the dielectric dimer was described by the temporal-coupled mode theory. A suitable model can be obtained by considering a single resonant mode coupled to three spherical harmonic scattering channels. The amplitude of the resonant mode $a(t)$ of the dimer driven by a quasi-monochromatic incoming field $\mathbf{s}^+(t)$ is governed by the usual dynamical CMT equation:

$$\frac{\partial a}{\partial t} = -i\left(\omega_1 - i\gamma_1\right)a(t) + \mathbf{K}\mathbf{s}^+(t) \qquad (S7.1)$$

where $\omega_1 - i\gamma_1$ is the eigenfrequency of the resonant mode of the dimer, and $\mathbf{K}$ is the vector of coupling constants of the mode to three scattering channels. The scattered field is given by the superposition of non-resonant (background) and resonant contributions:

$$\mathbf{s}^-(t) = \ddot{\mathbf{C}}\mathbf{s}^+(t) + \mathbf{D}a(t) \qquad (S7.2)$$

where $\ddot{\mathbf{C}} = \ddot{\mathbf{I}}$ describes the non-resonant background, and

$$\mathbf{D} = \mathbf{K} = \left(d_1, d_2, d_3\right)^T \qquad (S7.3)$$

Resolving the above equations in steady state with respect to $a$, we express the scattering matrix of the compact object with a single QNM:

$$\ddot{\mathbf{S}} = \ddot{\mathbf{C}} - i\frac{\mathbf{DK}}{\omega_1 - i\gamma_1 - \omega}, \qquad (S7.4)$$

Due to the coupling of the mode to a few scattering channels at once, the resulting S-matrix is clearly non-diagonal:

$$S_{\ell',\ell}^{q',q} \neq 0 \text{ for } \ell \neq \ell', q \neq q' \qquad (S7.5)$$

which is exactly what enables the multipole mixing.

Taking into account the double $m = \pm 1$ degeneracy of the mode, *partial* cross-sections in scattering channels $\left(q', \ell'\right)$ take the form:

$$\sigma_{q',\ell'} = \frac{2}{k^2}\left|\sum_{q,\ell}S_{\ell',\ell}^{q';q}\frac{c_\ell^q}{2} - \frac{c_{\ell'}^{q'}}{2}\right|^2 \qquad (S7.6)$$

where

$$c_\ell^{TE} = i^\ell\sqrt{\pi\left(2\ell+1\right)}, \quad c_\ell^{TM} = i^{(\ell-1)}\sqrt{\pi\left(2\ell+1\right)} \qquad (S7.7)$$

are the plane wave expansion amplitudes [4].

**The bibliography**

[1]   R. Alaee, C. Rockstuhl, and I. Fernandez-Corbaton, Exact Multipolar Decompositions with Applications in Nanophotonics, Adv Opt Mater 7, 1 (2019).

[2]   E. A. Gurvitz, K. S. Ladutenko, P. A. Dergachev, A. B. Evlyukhin, A. E. Miroshnichenko, and A. S. Shalin, The High-Order Toroidal Moments and Anapole States in All-Dielectric Photonics, Laser Photon Rev 13, 1800266 (2019).

[3]   A. K. Ospanova, A. Basharin, A. E. Miroshnichenko, and B. Luk'yanchuk, Generalized hybrid anapole modes in all-dielectric ellipsoid particles [Invited], Opt Mater Express 11, (2021).

[4]   J. D. Jackson, Classical Electrodynamics, 3rd ed. (John Wiley and Sons, Inc., New York, 1999)